# Openslice: An opensource OSS for Delivering Network Slice as a Service

Christos Tranoris
Electrical Eng. Dept.
University of Patras
Patras, Greece
tranoris@ece.upatras.gr, ORCID:
https://orcid.org/0000-0002-3433-037X

*Abstract*—As the 5G standards mature and awareness of the capabilities of the technology increases, industry verticals are becoming more eager to test new services and develop them to the level of maturity required for market adoption. Network slicing, i.e. multiple virtual networks running on a common infrastructure, is considered a key mechanism to serve the multitude of tenants (e.g. vertical industries) targeted by forthcoming fifth generation (5G) systems, in a flexible and cost-efficient manner. It is predicted that one of the most popular models for customers will be the Network Slice as a Service (NSaaS) model. This model allows a Network Service Customer to order and configure a Network Slice and offered it as a service. This work presents Openlice a Service based, opensource OSS for delivering NSaaS following emerging standards from SDOs. We strongly believe that such open source solutions make it easier for organizations to enable complex scenarios especially in the are of Non-Public Networks.

## I. Introduction

Forthcoming fifth generation (5G) systems are conceived as highly flexible and programmable end-to-end communication, networking, and computing infrastructures that are able to meet the diversified requirements from multiple services. Such services may require high multi-Gbps peak data speeds, ultra-low latency, massive device connectivity, reliability, increased network capacity, increased availability, and data-driven insights. At the same time, 5G systems are expected to operate in a flexible and cost-efficient manner, while also serving multiple tenants (e.g., verticals). The International Telecommunication Union (ITU) has categorized the services provided by 5G into major use cases [1]: enhanced mobile broadband (eMBB) including ultra-high definition (UHD) TV, massive machine-type communications (mMTC) for metering, logistics, smart agriculture, and ultra-reliable and low latency communications (URLLC) for autonomous driving and automated factory.

### A. Delivering Network Slice as a Service

Following the Anything-as-a-Service (XaaS) paradigm [2], strong attention has been drawn to the Network Slice-as-a-Service (NSaaS) model. NSaaS is more of a business-oriented concept than a technological one, able to map service demands automatically from a customer to functionalities, topology, policies, and parameters of a network slice, as well as to provide component-based and auto-configured network functionalities for operators to design and launch network services more conveniently. Some of the examples of NSaaS models include the Service Blueprints as defined in [3], [4] or the Generic Slice Template as defined by GSMA [5]. Under the NSaaS model, network slices are intended to be delivered to vertical customers as services. This model allows each customer to use the provided slice to set up and run one or more use cases with complete independence. An operator is allowed to compose and operate different network slices in parallel (e.g., to host multiple enterprises) while guaranteeing slice isolation so that data communication in one slice does not negatively affect services in other slices [2]. Based on Network Function Virtualization (NFV), network slicing realizes the service separation for multi-tenancy, so as to virtually build an exclusive network for each tenancy.

One of the main cornerstones of network slicing is network functions virtualization (NFV). ETSI NFV has defined an architectural framework [10], where the NFV technology contains the general-purpose processor platform, the cloud operating system, the hypervisor, distributed computing, and the software of network elements. It decouples software and hardware and shields the hardware details for virtual network functions (VNFs). More specifically, there is a management and orchestration (MANO) block, which controls the lifecycle of VNFs, as well as the connectivity among them. A customer orders a network slice from the OSS/BSS platform and that order is then passed onto the MANO platform to enable the brokering of NFV resources.

From the point of view of a vertical customer, the slice is seen as an isolated and tailored experimentation platform. However, an operator is allowed to compose and operate different network slices in parallel (e.g., to host multiple enterprises), as long as they guarantee slice isolation, so that data communication in one slice does not negatively affect services in other slices [8]. Each vertical needs to have only their own services exposed to them. This exposure can be defined as the ability of the facility service provider to securely expose the management capabilities of every slice instance towards the vertical customer. This requires proper isolation mechanisms to be embedded in the slice instance in order to avoid security and privacy breaches, say, between slice instances used by other vertical customers. By regulating the service exposure, the facility service provider can define the degree of visibility and control the vertical can have of the respective slice instance. In the 5G-VINNI white paper [9], the authors define different exposure levels in 5G-VINNI highlighting that higher the exposure level, the more advanced capabilities the vertical can get.

### B. Telecom industry organizations

The mission of these organizations is to collect information on service requirements from different vertical industry alliances (e.g. 5G-ACIA for Industry 4.0, 5GAA for automotive sector), identify potential technologies that can satisfy these requirements, and inform corresponding standards bodies, so that they can develop appropriate technology solutions. Relevant telecom industry organizations include Next Generation Mobile Networks (NGMN) Alliance, GSM Alliance (GSMA), Tele Management Forum (TM Forum) and Metro Ethernet Forum (MEF). In the context of the present paper, we will focus on GSMA and TM Forum.

On the one hand, GSMA is a trade body representing the interests of mobile operators worldwide. GSMA's work on network slicing focuses on how translating service requirements from industry use cases into network slice requirements, addressing the existing gap between vertical and telco industries. In order to offer verticals some guidelines on how to issue their service requirements towards operators, the GSMA has defined the Generic Network Slice Template (GST) [5]. The GST is a set of attributes that can be used to characterize any slice in terms of performance, functionality, operation and scalability. These attributes can be used by the operator and verticals to agree on an SLA. At this stage, 67 attributes have been defined in GST.

The GST provides a standardized list of attributes that can be used to characterise different types of network slice. GST is generic and is not tied to any: i) Type of network slice, ii) Agreement between a Network Slice Customer (NSC) and a Network Slice Provider (NSP). A Network Slice Type (NEST) is a GST filled with (ranges of) values. There may be two kinds of NESTs:

- Standardized NESTs (S-NEST), i.e. NESTs which character attributes are assigned (ranges of) values by SDOs, working groups, foras, etc. such as e.g. 3GPP, GSMA, 5GAA, 5G-ACIA, etc.;
- Private NESTs (P-NEST), i.e. NESTs which character attributes are assigned (ranges of) values by the Network Slice Providers, which are different from those assigned in S-NESTs.

Network Slice Providers can build their network slice product offering based on S-NESTs and/or their P-NESTs. For example, a Standardized Network Slice Type (S-NEST) [10]

TM Forum is actively working on the digital transformation and evolution of current Operations / Business Support Systems (OSS/BSS), seeking solutions that facilitate *i)* their consumption by verticals and *ii)* their integration into existing standards-defined architectural frameworks. In this respect, one of TM Forum's key contributions is the definition of the Open Digital Architecture [10], which is a multi-layer service platform that can be used by the operators to deliver XaaS, where X refers to the resource under consideration (e.g. infrastructure resource, network function, network service, etc.). To allow a given vertical to consume XaaS, the network operator provides him with corresponding resource management capabilities, offering them in the form of *Open APIs*. Defined in the TM Forum's Open API program, these APIs present two main features. First, they vendor-agnostic, which means they are not tied to vendor-specific management solutions. Secondly, they can be flexibly defined across layers; indeed, OpenAPIs at a given layer result from the composition of open APIs from the layer immediately below, following recursive patterns. These two features make OpenAPIs a good candidate for fast integration of components in multi-vendor environments, guaranteeing reusability and interoperability in the heterogenous 5G ecosystem.

The above are discussed in [11] and depicted in Figure 3. The NaaS API component suite supports the lifecycle functions required to manage the network capabilities exposed as Network as a Service and managed by operational domains. The TMFORUM defined component suite covers the operations required to be exposed in order to provide the functionality required by Operational Domains interworking with OSS/BSS applications and/or other domains from one service provider or from 3rd parties. One of the key requirements is the re-use of an API functionality rather than using a large set of specific APIs. This also simplifies the number of APIs needed and reduces the initial and maintenance costs as suppliers typically charge on a per API basis.

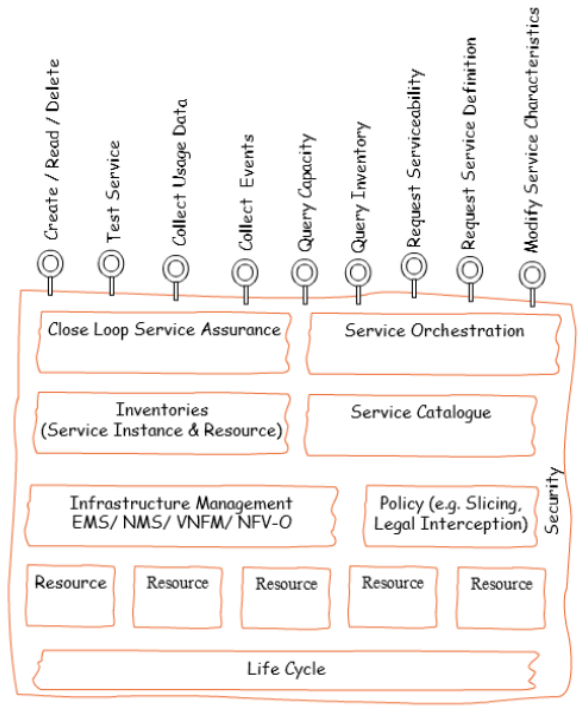

*Figure 1 Operational Domain Management Functions and Capabilities TMFORUM [4]*

C. *Standards Development Organizations*

Although there exists several standards bodies in telco ecosystem, two of them are key in the context of network slicing: the Third Generation Partnership Project (3GPP), focused on developing solutions and mechanisms for the support of slicing in mobile networks; and the European Telecommunications Standards Institute (ETSI), dealing with aspects related to the deployment of slices in virtualized environments.

*1) 3GPP*

3GPP consists of different working groups. One of them is SA5, in charge of specifying requirements, mechanisms and procedures for the provisioning of network slices in 5G mobile networks. This provisioning is done with a service-based management architecture [13], which executes the lifecycle of multiple network slices based on a well-defined information model [14]. This information model describes the functional components of the slice (network functions arranged into one or more network slice subnets) and allows specifying the service requirements it shall support. For the latter, the attributes defined in *ServiceProfile* struct are used. These attributes are translation of GST attributes into 3GPP domain.

3GPP SA5 and GSMA have collaborated in the complex work of designing a model-based network slice specification. This collaboration, executed through Liaison Statements (LSs) exchanged between both organizations, has been based on keeping Network Slice IOC definition aligned with the GSMA's Generic Slice Template (GST) specification. This alignment responds to the need for a consistent mapping of customer-facing service requirements (GSMA domain) to resource-facing service requirements (3GPP domain). It is built upon the idea that GST attributes representing network slice Service Level Specification (SLS) need to be translated into the 3GPP *ServiceProfile*. The *ServiceProfile* is a construct defined within the Network Slice IOC that allows for the service properties of a network slice (e.g. maximum/guaranteed supported downlink, maximum/guaranteed supported uplink, isolation, packet delay budget, etc.) to be defined.

The GSMA GST is used as the SLA information for the communication between the vertical industry and the communication service provider. The SLA requirements can be fulfilled from management aspect and control aspect in a coordinated way. The SLS includes ServiceProfile information model. [1] As shown in figure L.2.1, the GST [3] is translated and used as input to NRM ServiceProfile, the ServiceProfile can be translated to corresponding requirements for dedicated domains. For example, 5GC SliceProfile is used to carry 5GC domain requirements, NG-RAN SliceProfile is used to carry NG-RAN domain requirements, and TN requirements are translated and provide to TN domain. Some of the information in 5GC SliceProfile and NG-RAN SliceProfile translated to configurable parameters of network function for the control plane SLA support purpose.

Figure 2 shows how GST attributes are used by 3GPP as inputs to the *ServiceProfile* and then further translated into domain specific requirements. These requirements include 3GPP network slice subnet requirements, contained on individual *SliceProfile* (i.e. 5GC *SliceProfile* and NG-RAN *SliceProfile*), and TN requirements. Finally, these domain specific requirements are translated into domain specific configuration parameters, including 5GC, NG-RAN and TN configuration parameters. Some of these parameters may be injected on individual NFs (configurable parameters), while others will be kept at the management and orchestration level (non-configurable parameter). Examples of 5GC *SliceProfile* configurable parameters include "maxDlThroughputPerSlice" (triggers configuration of corresponding UPFs) and "maxNumberOfPDUSessions" (triggers SMF configuration). Examples of 5GC *SliceProfile* non-configurable parameters include "isolationLevel".

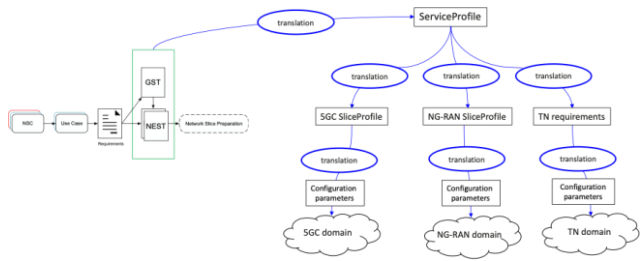

*Figure 2 The network slice journey - from GST to network slice configuration parameters (Source: [17])*

3GPP in [18] Annex A glues together GST and network slicing, in a so called Network Slicing journey (Figure 3). The journey describes the process of delivering a NSaaS from the Network Slice Customer order (Step 1), via Network Slice Provider's BSS/OSS (steps 2-6), until delivering and realizing the NS with VNF and/or PNFs (Step 7)

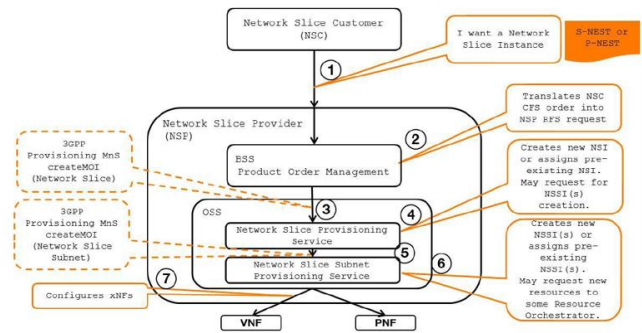

*Figure 3 3GPP Network Slice journey (NSaaS model) – high-level call flow*

*2) ETSI*

ETSI is formed of a wide variety of industry specification groups, among which NFV plays a key role. This group defines a management and orchestration stack that allows the deployment and operation of network services, each defined as a composition of one or more Virtualized Network Functions (VNFs) that can be flexibly allocated in different cloud sites. The core component of this stack is the NFV Orchestration (NFVO), which handles the lifecycle management of network services and their VNFs based on the information retrieved from their descriptors, namely Network Service Descriptors (NSDs) [6] and VNF Descriptors (VNFDs) [6].

The relationship between 3GPP and NFV is as follows: the virtualized part of a slice can be deployed as a network service. This means mapping *ServiceProfile* into the corresponding NSD that NFVO can consume and understand. For communication between 3GPP and NFV, the NFVO offers a northbound interface [21] with multiple capabilities, including NSD management as well as network service lifecycle, performance and fault management.

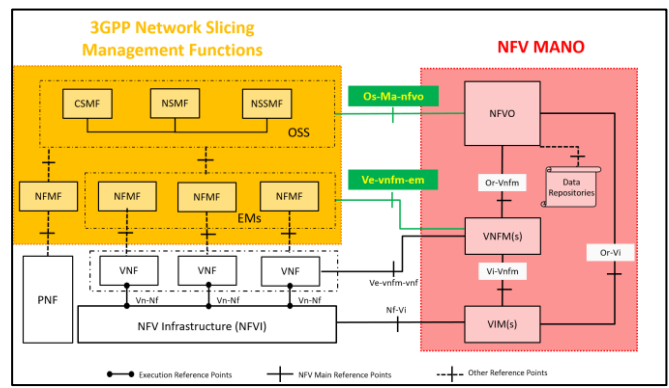

*Figure 4 3GPP network slicing management in NFV framework (adapted from [22] [23])*

As seen from the above-referred description, 3GPP can handle all the activities that go beyond virtualization, and hence are out of scope of NFV MANO. These activities include (i) the deployment and operation of PNFs, (ii) VNF application layer configuration and management, and (iii) application-aware network service configuration and management. To carry out these activities, 3GPP network slicing management functions need to consume data and operations from the reference points exposed by NFV MANO stack, so an interaction between two frameworks are required. Taking into account the relationship between a 3GPP network slice and a network service as discussed earlier, ETSI NFV-EVE 012 [23] proposes in Figure 4 a first approach on how the intended interaction could take place, showing the reference points that can be potentially (re)used for this end: Ve-vnfm-em and Os-Ma-Nfvo [21].

## II. APPROACH

### A. Requirements for delivering NSaaS

If one adopts NSaaS as a service delivery model, it means the the entire Network Slice Provider provisions end-to-end network slices to verticals customers upon request. Every vertical makes use of the provided slice to onboard their use cases. To ensure reproducibility (ability to replicate tests across different sites to assess KPIs under different load conditions) and interoperability (ability to ensure the interworking between the tools deployed on each site, typically from different vendors), the proposed architecture incorporates the guidelines form telecom industry organizations, and the normative specifications from standards bodies. This means that:

- The Network Slice Provider offers TM Forum Open APIs, making them available for verticals, so they can invoke relevant NSaaS operations.

- The Service Orchestrator implements the 3GPP network slice management functionality, taking the roles of both Network Slice Management Function (NSMF) and Network Slice Subnet Management Function (NSSMF).

- The Network Orchestrator is interconnected with a NFVO (NFV Orchestrator), which orchestrates instances of VNFs and network services with the help of a Virtualized Infrastructure Manager (VIM). The VIM manages the virtual resources on top of which those instances are executed.

The NFVO offers the northbound interface [24] towards the Service Orchestrator using ETSI SOL005 [21]. This normative specification defines the protocol and data model for the interface capabilities, in the form of RESTful APIs. These APIs have been become de-facto solutions for most of industry and open-source NFVOs.

### B. Openslice and Delivering a NSaaS: From Service order to E2E Service

Given the status described in previous section, we started designing and implementing a service-based architecture which utilizes well-known standards and delivers Network Slices as Services. Figure 5 depicts how Openslice is positioned in the ETSI NFV architecture.

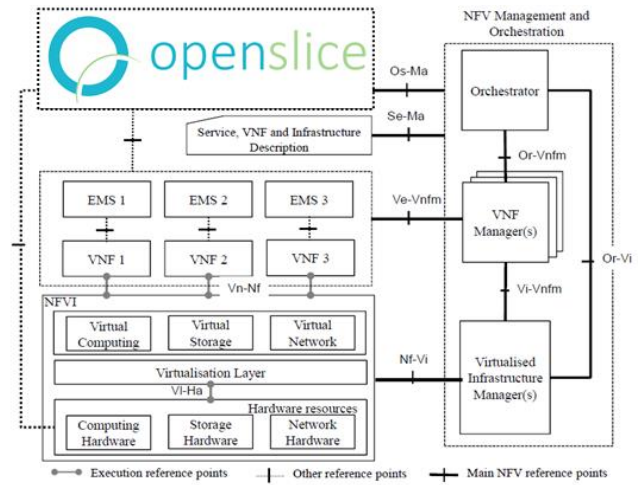

Figure 5 Positioning Openslice in ETSI NFV architecture

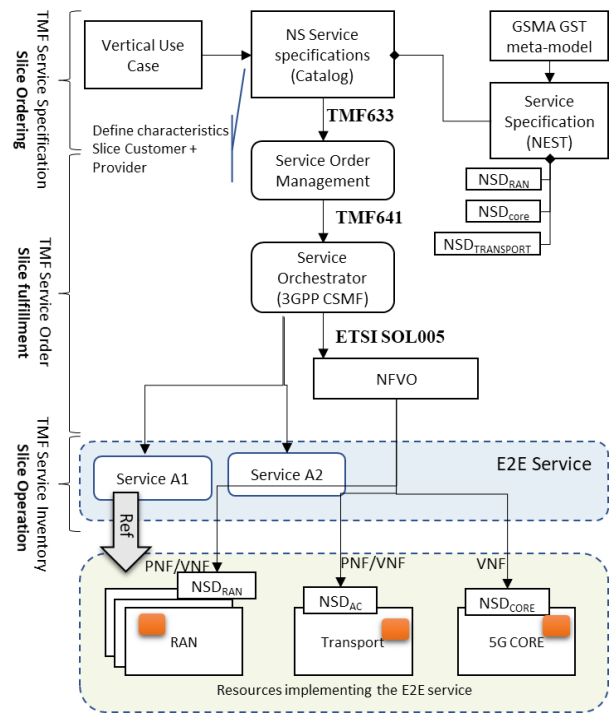

Figure 6 Network Slice Instantiation in Openslice

Figure 6 displays how Openslice approaches the NSaaS Instantiation and delivery of an E2E service, i.e. from radio to the core involving VNFs/PNFs. There are three main phases Slice Ordering, fulfillment, and Operation.

#### 1) Slice Ordering

It is assumed that Slice Ordering is initiated from a 5G Vertical (a NS Customer), by ordering specific Service Specifications that are related with Network Services. These Service Specifications are located in a Service Catalog of the provider. Both Specification and Service Catalog follow the model of (**TMF633**), which provides artifacts (e.g. models and dependencies) for the specification of Services. We have two types of Service Specifications: Customer Facing Services (CFS) and Resource Facing Services (RFS). CFSs are exposed to the catalog and users can order them. RFSs are internal and related with the underlying resources. Assuming these service specifications follow the GST model, the derived

NESTs are modeled as TMF Service Specifications. Section III describes how we created a GST meta-model in TMF model.

A NEST (a Service Specification) can consist of several other services, so we call it a **Service Bundle**. It consists of RFSs having each RFS related with underlying NSDs(VNFs) for radio, core, or other applications. For example, a service spec for a 5G Slice on Country Greece and Location Athens, might consists of RFSs that will configure via VNFs and PNFs, RANs in Athens and create a new 5GCore while configuring an existing transport network.

Thus, the Customer requests the Service Spec and creates a Service Order. Service Orders are modeled with the **Service Ordering API (TMF641),** which allows issuing a service order, and includes selected Service Specifications as well as instantiation parameters. (e.g. Network Slice region, QoS, Number of UEs, etc )

*2) Slice fulfillment*

The **slice fulfillment phase** begins when the customer triggers a service order from the service catalog. In this service order, the customer provides a completed specification of the slice instance he wants, including information on slice topology (possibly extended with 3$^{rd}$ party VNFs) and slice attributes (filled in with values fitting use case requirements). To do that, the vertical makes use of the Service Ordering API. The issued service order is then captured by the Service Order Management (SOM), which propagates it towards the corresponding g Service Orchestrator(s) again via TMF641 API calls. Next, the service order is processed.

This processing consists in translating the received service order (CFS, handled by the Service Orchestrator) into a set of resource requirements for the network slice to be instantiated (RFS, handled by the NFVO). To successfully achieve this translation, Service Orchestrator and NFVO exchange information relying on ETSI SOL005 capabilities. Once this translation is completed, the NFVO allocates the slice instance, instructing the VIM for that end.

At this stage we have the E2E network service running involving Radio, 5G Core, Transport or any other desired Network Service (e.g. a firewall).

*3) Slice Operation*

As described later, TM Forum OpenAPIs may allow customers not only to interact with the service catalog but also to consume the exposed capabilities.

In the **slice operation phase,** the slice is already instantiated, and can be made available for operation. During this phase, the customer keeps track of the status of the slice instance, making use of the Service Inventory API **(TMF638),** which defines standardized mechanisms for CRUD operations over the records providing run-time information about deployed slice instances.

III. GENERIC SLICE TEMPLATE TRANSFORMED INTO TMF633 SERVICE SPECIFICATION

This section provides the process of transforming the GST specification as defined by GSMA v2.0 to the Service Specification entity model as defined by TMFORUM in TMF633 Service Catalog API (release 18.5.0)

GST will be modeled as a Service Specification entity and will have some default values. The idea though will be to "clone" the Service Specification and create a specific one with certain values (i.e. create a NEST – Network Slice Type)

The latest version of the GST modeled as json format following the TMF633 Service Specification entity is maintained at: https://github.com/openslice/io.openslice.tmf.api/blob/master/src/main/resources/gst.json

*A. Approach*
   *1) GST Attribute name*

Each GST Attribute e.g. "Area of Service" will be of type **serviceSpecCharacteristic**.

*2) Allow the customer to configure value*

**configurable**: if True means that the end user can configure it during a Service Order

*3) Generic properties*

**description**: The description as given in the GST

*4) Attribute presence*

Attribute presence is related to **minCardinality**.

- If in GST is Mandatory then "minCardinality": 1
- If in GST is Conditional then "minCardinality": 0
- If in GST is Optional then "minCardinality": 0

*5) Value*

**valueType**: is a String both in GST and in Service Specification. See next section for possible values

if SET or ARRAY then the type of the elements is given by the **serviceSpecCharacteristicValue**

Measurement unit in GST is **unitOfMeasure** in **serviceSpecCharacteristicValue.**

*6) Approach for Composite attributes*

There are many attributes in GST that are "Composite". For example:

- **Area of Service** attribute includes also the **Region specification**.
- **Deterministic communication** attribute includes the **Availability** attribute and the **Periodicity** attribute.
- **Isolation level** attribute includes the **Isolation, Physical Isolation** and **Logical Isolation attributes**
- etc..

For attributes like the above, we use the **serviceSpecCharRelationship** property of the Service Specification. To indicate the relationship, we add an entity with

- role: the name of the part (e.g. **Region specification**)

- name: The name of the part as it will be found later in the Service Specification (e.g. **Area of Service: Region specification**)
- **relationshipType=dependency**

Examples:

For **Area of Service** the **serviceSpecCharacteristic** has in **serviceSpecCharRelationship**:

```
"serviceSpecCharRelationship": [
    {
        "role": "Region specification",
        "name": "Area of Service: Region specification",
        "relationshipType": "dependency"
    }
```

For **Energy efficiency** the **serviceSpecCharacteristic** has in **serviceSpecCharRelationship**

```
"serviceSpecCharRelationship": [
    {
        "role": "Network slice energy efficiency",
        "name": "Energy efficiency: Network slice energy efficiency",
        "relationshipType": "dependency"
    },
    {
        "role": "Time frame of the measurement",
        "name": "Energy efficiency: Time frame of the measurement",
        "relationshipType": "dependency"
    }
]
```

*7) GST Conditionals*

Conditionals: we write to regex some conditions when an attribute is *Conditional*. Examples:

"regex": "Conditional: This parameter must be present when "Isolation" is set to 1.",

*8) Tags*

Useful in future to provide categorization. Tags are used as labels attached to the attributes to give additional information about the nature of each attribute.

To define the tags we use the **serviceSpecCharRelationship** property of the Service Specification.

"role": "tag", "relationshipType": "tag" are used in serviceSpecCharRelationship:

Examples

```
"serviceSpecCharacteristic": [
{
   "name": "Area of Service",
   "configurable": false,
   "description": "This attribute specifies the area where the terminals can access a particular network slice. Therefore, the attribute specifies the list of the countries where the service will be provided. The list is specific to NSPs and their roaming agreements",
   "extensible": null,
   "isUnique": true,
   "maxCardinality": 1,
   "minCardinality": 0,
   "regex": null,
   "valueType": "SET",
   "serviceSpecCharRelationship": [
       { "name": "Character Attribute", "role": "tag", "relationshipType": "tag" },
       { "name": "Operational", "role": "tag", "relationshipType": "tag" },
       { "name": "Scalability Attribute", "role": "tag", "relationshipType": "tag" },
       { "name": "KPI", "role": "tag", "relationshipType": "tag" } ,
       {
           "role": "Region specification",
           "name": "Area of Service: Region specification",
           "relationshipType": "dependency"
       }
   ],
```

*B. General considerations*

*1) **valueType** is a String.*

The allowed values are:

> `INTEGER`,
> `SMALLINT`,
> `LONGINT`,
> `FLOAT`,
> `BINARY`,
> `ARRAY`,
> `SET`,
> `BOOLEAN,`
> `TEXT`
> `LONGTEXT`
> `ENUM`,
> `TIMESTAMP`;

*2) A value is of type ANY in the model.*

We have defined ANY as an object with two properties:

*Value* which is a string and *alias* which is used to describe a value (useful in enums)

```
"value": {
       "value": "0",
       "alias": "Not allowed"
   }
```

*C. Deviations*

*1) Area of Service: Region specification*

We made it a SET of INTEGER. GST has it as an INTEGER

*2) Slice quality of service parameters: Packet Error Rate*

Packet Error Rate to FLOAT instead of integer

*3) Uplink throughput per network slice: Guaranteed uplink throughput*

To bps instead of Bytes

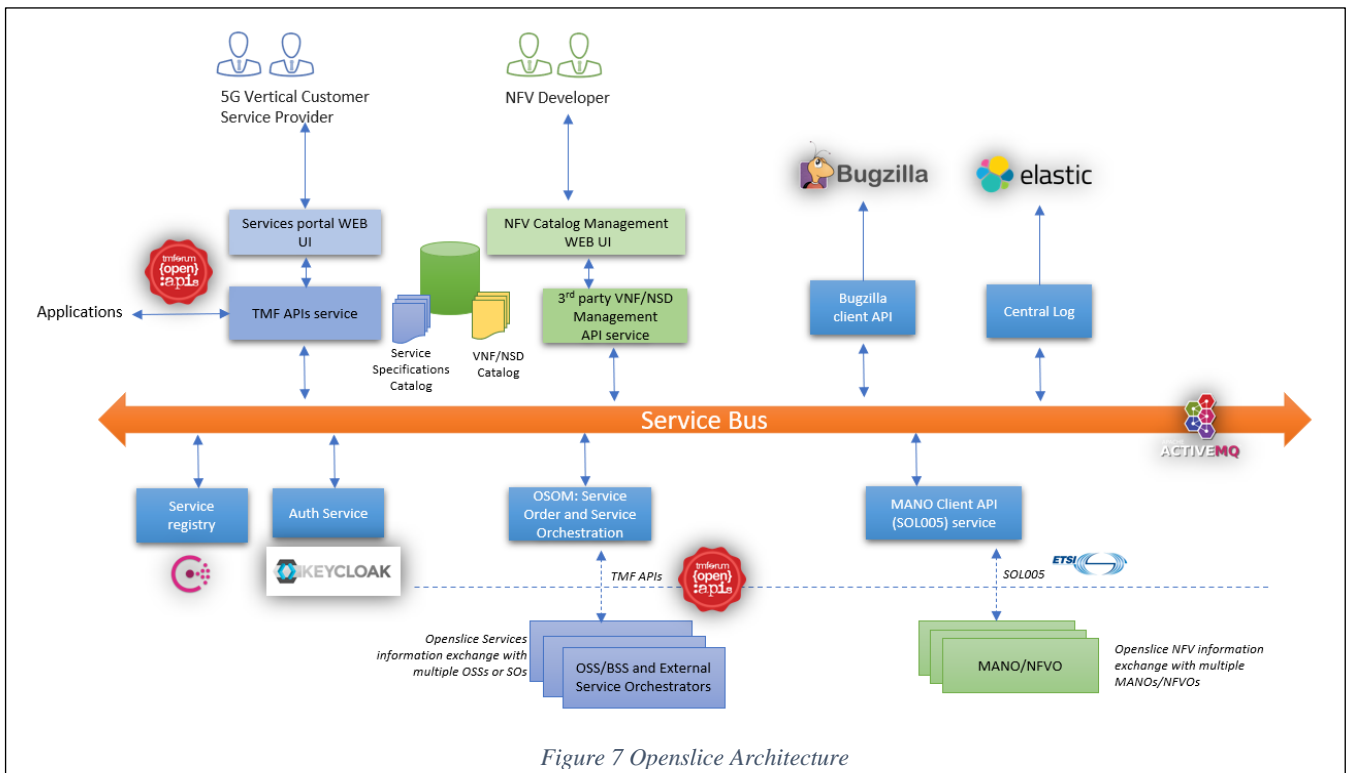

*Figure 7 Openslice Architecture*

## IV. OPENSLICE ARCHITECTURE

Openslice allows Vertical Customers to browse the available offered service specifications and also allows NFV developers to onboard and manage VNF and Network Service artifacts. There are two portals offering UI friendly access to users:

- The Services portal allows users to access services and service providers to design services.
- The NFV portal allows users to self-manage NFV artifacts and onboard them to a target MANO/NFV Orchestrator.

3rd party applications can use Openslice through TMForum Open APIs. In general Openslice offers the following main functionalities:

- Service Catalog Management: A CSP will have the ability to manage the Service Catalog Items, their attributes, organize in categories and decide what to make available to Customers
- Services Specifications: A CSP will be able to manage Service Specifications
- Service Catalog Exposure: A CSP will be able to expose catalog to customers and related parties
- Service Catalog to Service Catalog: Openslice able to consume and provide Service Catalog items to other catalogs
- Service Order: The Customer will be able to place a Service Order
- Service Inventory: The Customer and Provider will be able to view deployed Services status

Figure 7 Openslice ArchitectureFigure 7 displays the overall architecture of Openslice. Openslice allows Vertical Customers browsing the available offered service specifications. It consists of:

- Web frontend UIs that consist of mainly two portals: i) a NFV portal allowing users self-service management and onboarding VNFDs/NSDs to facility's NFVO ii) a Services Portal, which allows users to browse the Service Catalog, Service Blueprints specifications and the Service Inventory
- An API gateway that proxies the internal APIs and used by the web front end as well as any other 3rd party service
- A Message Bus where all microservices use it to exchange messages either via message queues or via publish/subscribe topics
- An authentication server implementing Oauth2 authentication scheme
- A microservice offering TMF compliant API services (eg Service Catalog API, Service Ordering APIetc)
- A microservice offering NFV API services (eg VNF/NSD onboarding etc) and allows to store VNFDs and NSDs in a catalog
- A microservice that is capable to interface to an issue management system. For example it raises an issue to all related stakeholders (CSP, NOP, CSC) that a new Service Order is requested
- Central logging microservice that is capable to log all distributed actions in to an Elasticsearch cluster
- A Service Orchestrator solution that will propagate Service Ordering requests to the equivalent SOs and NFVOs

Next table provides a list of technologies.

TABLE I. OPENSLICE MICROSERVICES

| μService | Description | Technology |
|---|---|---|
| Auth | Server to authenticate/authorize verticals via OAuth 2.0 | Keycloak |
| metrics | Distributed tracing system that is used to troubleshoot latency problems in μservice architectures. | Zipkin |
| μService Registry | One stop solution for typical procedures in μservice architectures, including service (self) registration, discovery, key-value store and load balancing. | Consul |
| Issue mgmt client proxy | Offers interface to Bugzilla, which is a ticketing tool that allows issue tracking (fault alarms, service orders) and reporting (to verticals, facility operators, etc.) via tickets. | Java |
| Central Logging | Logs all distributed actions into an ELK (Elasticsearck, Logstash and Kibana) stack. | Java |
| Services OpenAPIs | Offers TM Forum's OpenAPIs to allow consumption of service catalogue exposed capabilities. These open APIs include Service Catalog, Ordering and Inventory APIs. | Java |
| NFV Mgmt APIs | Offers NFV APIs to manage 3rd party VNFDs (e.g. on-boarding, updating). These APIs allow verticals to bring their own VNFs to 5G-VINNI, to validate their KPIs. | Java |
| SOM and Orchestration | Referred as to OSOM, captures service orders triggered by verticals and propagate them to the corresponding SO. In case a service needs to be deployed across two (or more) sites, the SOM shall first decompose the received order. | Open source Engine Flowable |
| Messaging Bus | It allows the microservices to exchange messages, either in Publish/Subscribe scheme or via Messagin Queues | ActiveMQ |

A. *The Service Catalog*

The Service Catalog offers various ways to access the defined Service Specifications, organized in categories (e.g. eMBB, Networking, edge, etc). Customers can browse the publicly available catalog (Figure 9) and see offered services. (Examples from [25])

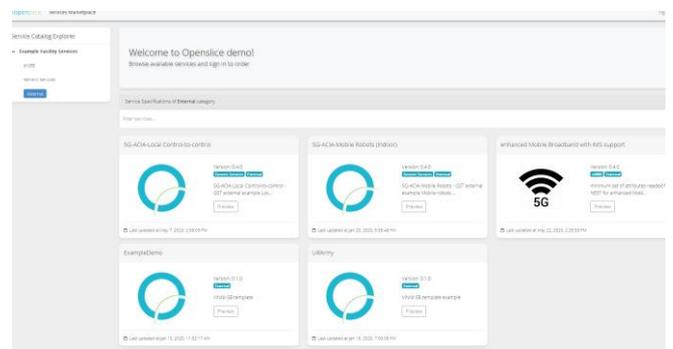

Figure 9 Openslice Service Catalog example of CFSs

Details of each service are available, so that the customer can understand underlying details (see Figure 10), as well as the underlying attributes of the service offered. Also any user defined attributes as well as ways to access the final operation service are also available. All this information is made from the Service Specification design.

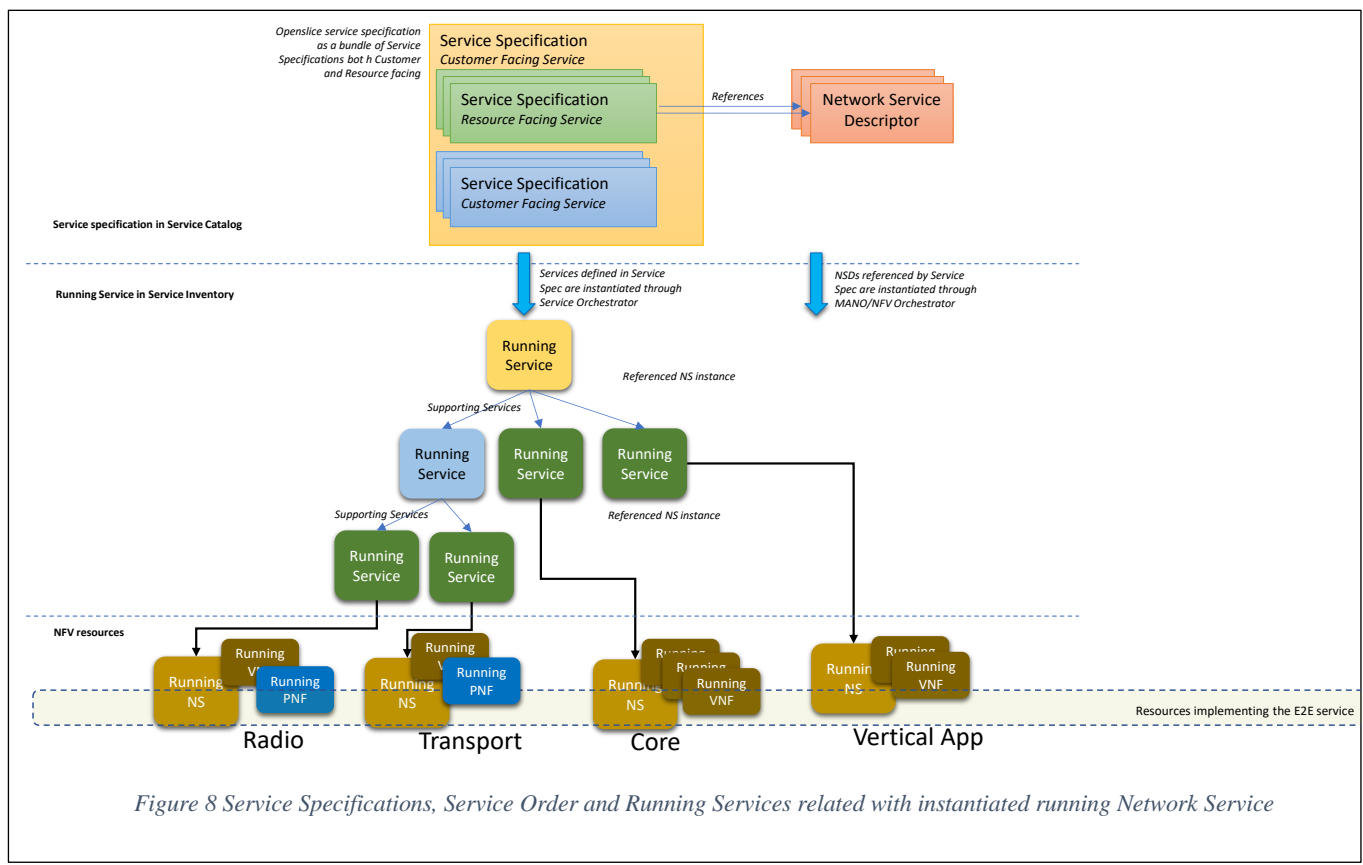

Figure 8 Service Specifications, Service Order and Running Services related with instantiated running Network Service

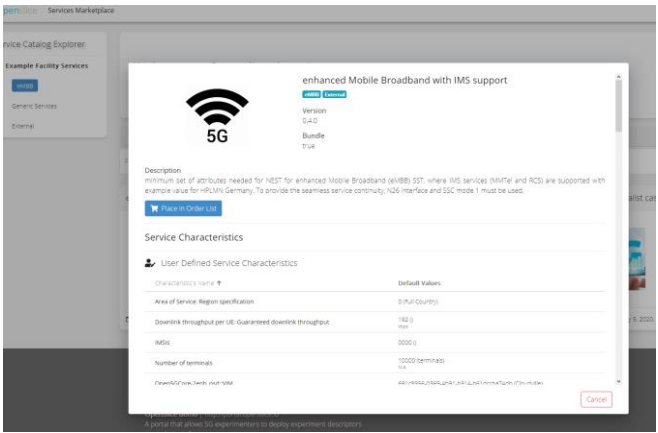

*Figure 10 Openslice Service CFS example*

Figure 11 displays a list of available CFSs and RFSs in the system. This is available to the service developers and admins. They can use it to construct new Services as bundles: Services that are related with other services.

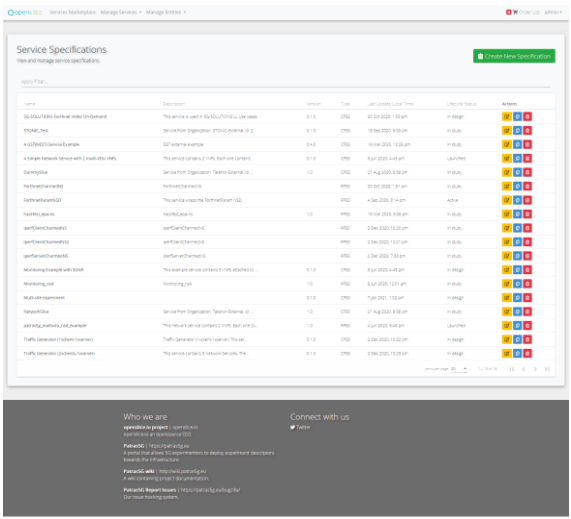

*Figure 11 List of Service Specifications, both RFSs and CFs*

Figure 12 depicts an example of a Service Specification Design. Provided that Openslice has the underlying model from TMF the users can either select to design their Service based on a GST or from scratch.

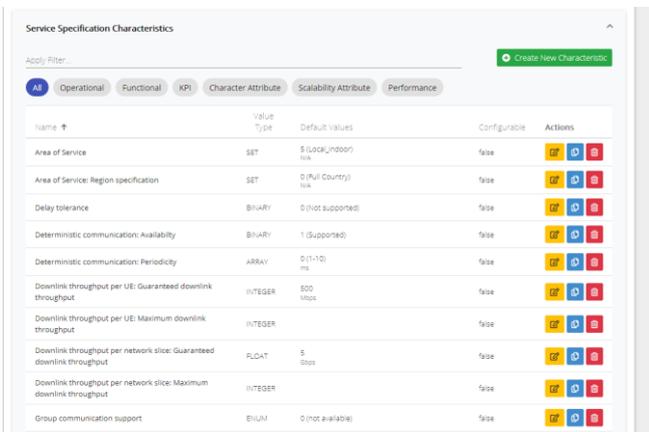

*Figure 12 Openslice Service Specification design example*

### B. The NFV Services

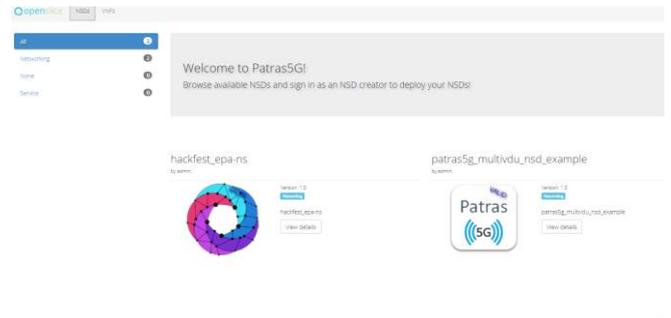

*Figure 13 The NGV Catalog*

The NFV services provide to the operator a way to onboard NFV artifacts. These NFV artifacts are then onboarded to the interconnected NFVOs of Openslice.

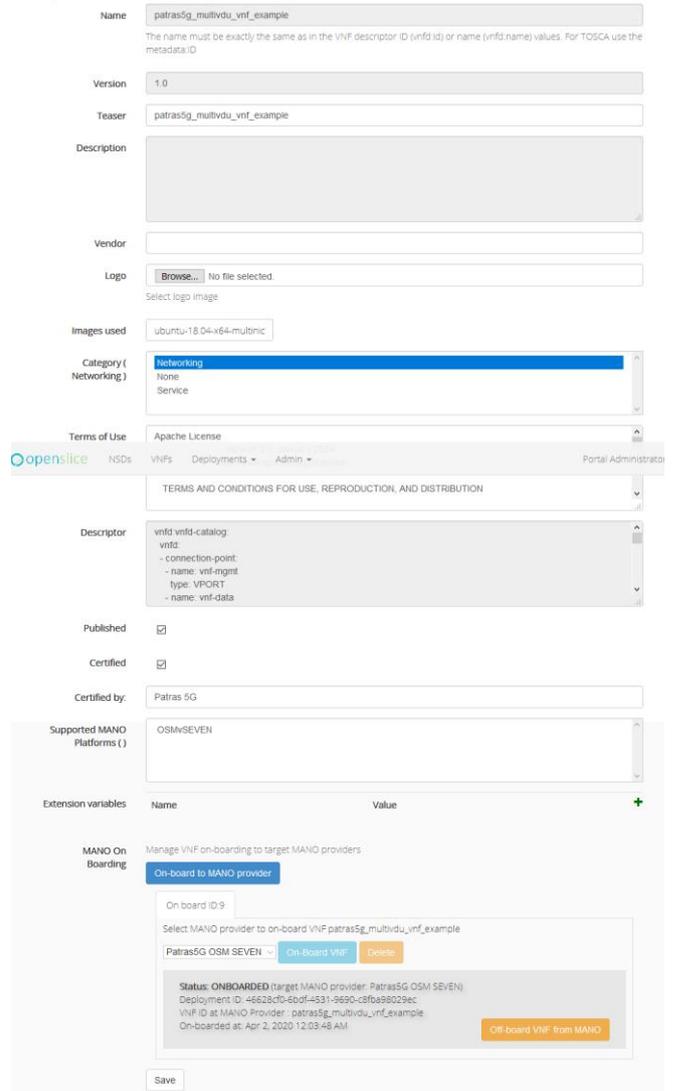

*Figure 14 Edit an Onboarded VNF*

For example Figure 13 displays available onboarded VNFs. Figure 14 displays an example VNF and the status of this onboarded to target NFVO. In our case to OSM SEVEN

## C. The running Network Service

Customers make Service Orders and Openslice instantiates the requested Service Specifications of the Service Order. Running Services instantiated by Openslice, reside in Openslice Service Inventory. The following picture displays how Service Specifications are related to Running Services and how Running Services relate with instantiated running Network Service.

Figure 8 depicts how Service Specifications, Service Order and Running Services related with instantiated running Network Service. As discussed in Section II a Service Specification can consist of several other services, called as a Service Bundle. It consists of RFSs having each RFS related with underlying NSDs.

The Service specifications defined in the Service Order are instantiated through the Service Orchestrator. The NSDs referenced by each RFS Specification are instantiated through MANO/NFV Orchestrator. So, there is a top-level Service instantiated reflecting the user service, e.g. eMBB 5G slice.

This is decomposed and references 3 other Services which run and registered in Openslice Service Inventory. Other Running Services maybe further decomposed to other Services. This is done until each service is mapped to exactly one Running NS in the underlying Infrastructure. The Running NS is implemented by the VNFs and PNFs of the Communication Service Provider.

## V. SERVICE ORDERING MANAGEMENT

Service Specifications reside into Service Catalogs, grouped in Categories. Openslice offers a Service Orchestrator called OSOM. OSOM instantiates Service Specifications by requesting Network Services from target MANOs/NFVOs. OSOM is a service responsible for:

- Service Order Management (SOM)
- Service Orchestration (SO)

It uses open source Flowable Business process engine (https://www.flowable.org). A Service Order follows the states as defined in TMF641 specification

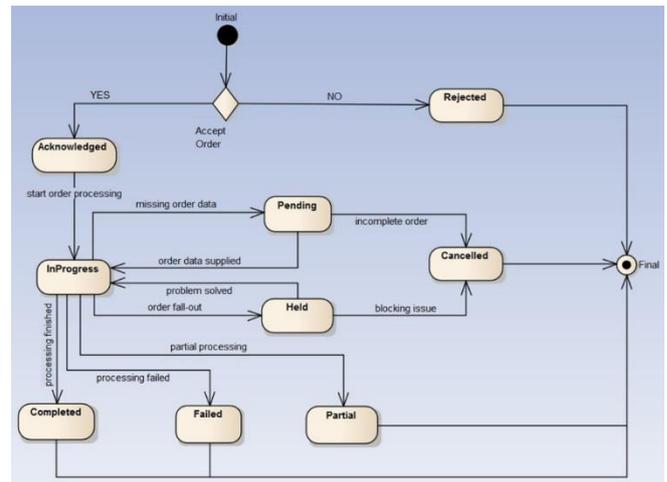

Figure 15 Service Order follows the states as defined in TMF641 specification

OSOM contains many processes for different orchestration cases. For example the Order Fulfillment is like the following figure expressed in BPMN. The BPMN diagram has an XML equivalent which is consumed by the process engine and executes the orchestration.

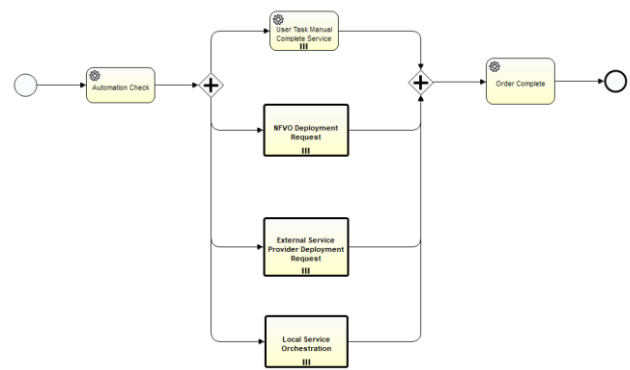

Figure 16 Order Fulfillment process

Here is the process logic:

The process first checks which tasks can be orchestrated by NFVO, which by human and which must be send to another partner external domain. We create at this stage the equivalent Services (TMF638/640 model) for this order O1.

1. During check it should decide to create Service(s) for this service order O1 and send it to ServiceInventory
2. The Services(TMF638 model) are assigned to the Order O1 In supportService List
3. Each OrderItem OI1 is related to one serviceSpecification
4. Each ServiceSpecification has also related serviceSpecRelationships
5. So if we receive an order O1 for a ServiceSpec A which relates to (a bundle of) 3 specs(2 CFS, 1 RFS) we do the following:
    5.1. Create a Service S_A(TMF638 model) for ServiceSpec A for Order O1
    5.2. We create also 3 Services S_C1, S_C2 and S_R1 equivalent to the serviceSpecRelationships (2 CFS, 1 RFS)

5.3. At this point the order will have 1 Local Service Orchestration Process(S_A), 2 supportingServices refs(S_C1, S_C2) and 1 supportingResource(S_R1)
5.4. The 3 supportingServices and 1 supportingResource correspond to total 4 Services in ServiceInventory
5.5. Service S_A will have:
   5.5.1. startMode 1: Automatically by the managed environment
   5.5.2. State: RESERVED and the Lifecycle will be handled by OSOM
6. Services S_C1 and S_C2 we decide that cannot be orchestrated then they have
   6.1. startMode: 3: Manually by the Provider of the Service
   6.2. State: RESERVED and the Lifecycle will be handled by OSOM
   6.3. If the CFS is a bundle spec it is further recursively orchestrated
7. Service S_R1 will have
   7.1. startMode 1: Automatically by the managed environment.
   7.2. State: RESERVED
   7.3. IF The Service has the characteristic CharacteristicByName( "NSDID") it will be further processed by the NFVO

There will be two instances of task "User Task Manual Complete Service" each for Services S_C1 and S_C2. The task is Transient for now. It displays only the services that are not automated! Here is a flow for future:

1. We wait here for human decision.
2. From API we get a result: a. If set to ACTIVE/TERMINATED then we complete the task b. In any other state we stay in this task until it is resolved as in step a. c. The Status of ORDER O1 is also updated to PARTIAL

There will be an instance of NFVODeploymentRequest process each for Service S_R1. (see later)

1. This process is related with the NFVO orchestration
2. It will send a msg to NFVO for a specific deployment request

All services in "Order Complete" are in a status:

1. Depending on the result the service S_A is either ACTIVE or INACTIVE or TERMINATED
2. The Status of ORDER O1 is also updated to COMPLETED or PARTIAL (in case we have some services running) or FAILED (in cases we have errors)

A Service follows the states as defined in TMF638 Service Inventory specification as show in next Figure 17.

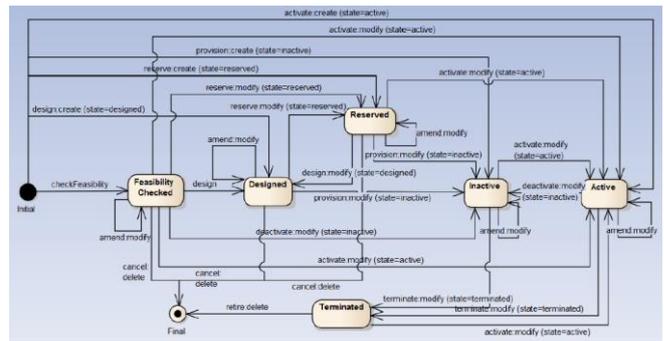

*Figure 17 TMF638 Service Inventory state diagram*

*Figure 18 Service Order Overview*

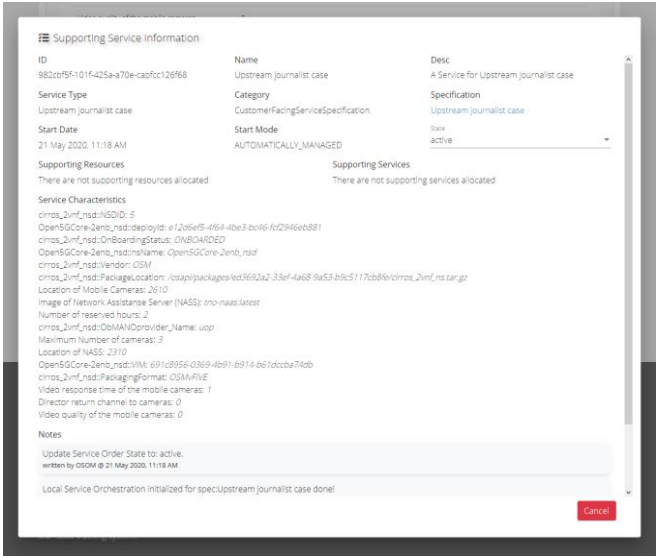

*Figure 19 Service Inventory Information*

Figure 18 displays a Service Order, while it is fulfilled by the OSOM. User is notified offline via automated emails by the issue management system when the service is ready. Figure 19 displays the running service with information from the service inventory. For example IP addresses or other service status is available here.

## VI. MULTIDOMAIN SCENARIOS AND FEDERATION

Openslice can be used to exchange service specifications/catalogs and make service orders between Organizations. A typical deployment across domains, involves today some typical components: i) an OSS/BSS to allow customers access the service catalog and perform service orders, ii) a Service Orchestrator (SO) component for executing the service order workflow, as well as iii) a Network Functions Virtualization Orchestrator (NFVO) for configuring the iv) network resources.

TMF Open APIs are introduced not only for exposing catalogues and accepting service orders, but also implementing the East-West interfaces between the domains, fulfilling also the LSO requirements as introduced by MEF. The following Figure 20 shows how Openslice could be used in such scenarios:

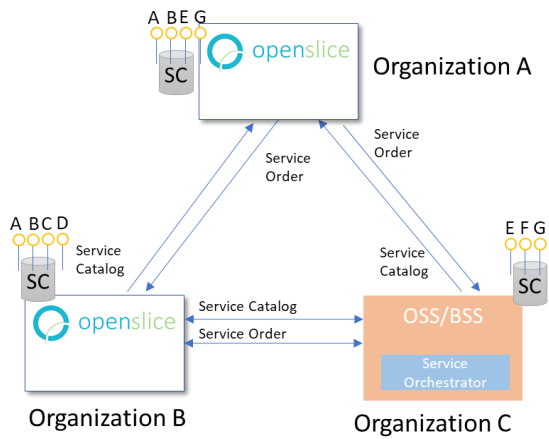

*Figure 20 Openslice in a mesh connectivity for service exchange between Providers*

In Openslice we can consume services from 3rd parties via Open APIs. We use the TMF 632 Party Management model to specify Organizations that we can exchange items and other information such as:

- Import Service Specifications
- Create a Service Order
- Use the Service Inventory to query the status of the service ordered to the external partner organization

## VII. CONCLUSIONS AND FUTURE WORK

Openslice is an Open Source OSS for the emerging 5G era. It follows telco standards and the NSaaS model make it ideal for interactions with other external services and parties.

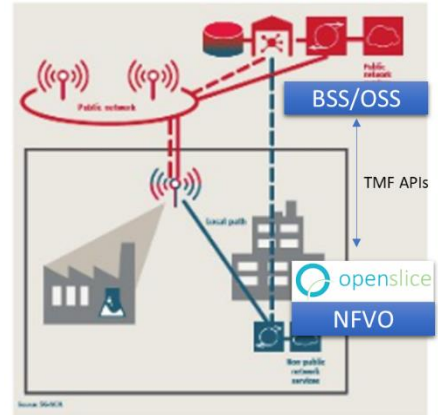

*Figure 21 Openslice in NPN scenarios*

For example as Figure 21 displays, Openslice can be used by Organizations, SME etc on Non-Public Networks, in case they would like to create and deploy services on campus, or in certain places and they would like also to interact with a public mobile network to request e.g. a shared slice.

Supported TMF APIs in Openslice as of today are the following:

- TMF 620 - Product Catalog Management
- TMF 622 - Product Ordering Management
- TMF 633 - Service Catalog Management
- TMF 634 - Resource Catalog Management
- TMF 638 - Service Inventory Management
- TMF 640 - Service Activation and Configuration
- TMF 641 - Service Ordering Management
- TMF 666 - Account Management specification
- TMF 669 - Party Role Management
- TMF 629 - Customer Management
- TMF 632 - Party management
- TMF 691 - Federated identity

Some of our next steps are the following:

- 3GPP CSMF Compliance (Communication Service Management Function)
- Service Assurance, Alarms and Events handling
- Mapping of user defined service characteristics to Network Service requirements

- Full Lifecycle management of Services (Network Slices) – modifications while service is active. This is available today as Day2 configurations in NFVO

*A. Public access*

**Documentation**: http://openslice.io

**Source code**: https://github.com/openslice

**Openslice demo**: http://portal.openslice.io/

**Supported APIs**: http://portal.openslice.io/tmf-api/swagger-ui.html

ACKNOWLEDGMENTS

This work is supported by the H2020 European Projects 5G-VINNI (grant agreement No. 815279), 5G-VICTORI (grant agreement No. 857201), 5G-SOLUTIONS (grant agreement No. 856691), and HELNET (MIS 5002781) co-financed project by Greece and the EU.